\documentclass[fleqn,usenatbib]{mnras}
\usepackage[normalem]{ulem} %this package can be removed
\usepackage{newtxtext,newtxmath}
\usepackage{ulem}

\usepackage[T1]{fontenc}
\usepackage{ae,aecompl}
\usepackage{comment}

\usepackage{stackengine,graphicx}	% Including figure files
\usepackage{amsmath}	% Advanced maths commands
\usepackage{amssymb}	% Extra maths symbols

\usepackage{multirow} 
\usepackage{multicol}

\newcommand*{\mysim}{\mathord{\sim}}

\newcommand{\nus}{\textit{NuSTAR}}
\newcommand{\chan}{\textit{Chandra}}

\newcommand\rxs{\mbox{1RXS~J170849.0--400910}}
\newcommand\rx{RX~J1713}
\newcommand\rxf{\mbox{RX~J1713.7--3946}}
\newcommand\wga{\mbox{1WGA~J1713.4--3949}}
\newcommand\xmm{\textit{XMM-Newton}}
\newcommand\mos{\textit{MOS1+2}}
\newcommand\ib{\textit{IBIS}}
\newcommand\intl{\textit{INTEGRAL}}
\newcommand\rosat{\textit{ROSAT}}

\title[SNR \rx\ spectrum steepening from XMM to INTEGRAL]{Progressive steepening of the SNR \rxf\ X-ray spectrum from {\it XMM-Newton} to {\it INTEGRAL}}

\author[Kuznetsova et al.]{Ekaterina Kuznetsova$,^{1}$\thanks{E-mail: eakuznetsova@cosmos.ru}
Roman Krivonos$,^{1}$ Eugene Churazov$,^{1,2}$ Natalia Lyskova$,^{3,1,4}$ \and
Alexander Lutovinov$^{1,3}$
\\
$^{1}$Space Research Institute of the Russian Academy of Sciences, Profsoyuznaya 84/32, 117997 Moscow, Russia\\
$^{2}$Max-Planck-Institut f\"ur Astrophysik, Karl-Schwarzschild-Strasse 1, 85741
Garching, Germany\\
$^{3}$National Research University Higher School of Economics, Myasnitskaya str. 
  20, Moscow 101000, Russia\\
$^{4}$ ASC of P.N.Lebedev Physical Institute, Leninskiy prospect 53, Moscow 119991, Russia \\
}

\date{Accepted XXX. Received YYY; in original form ZZZ}
\pubyear{2019}

\begin{document}
\label{firstpage}
\pagerange{\pageref{firstpage}--\pageref{lastpage}}
\maketitle

%%%%%%%%%%%%%%%%%%%%%%%%%%%%%%%%%%%%%%%%%%%%%%%%

\begin{abstract}

In this work, we present the first detailed analysis of the supernova remnant \rxf\ in the hard X-ray energy range with the \ib\ coded-mask telescope on board the INTEGRAL observatory. The shell-type morphology of the entire remnant is mapped in hard X-rays for the first time and significantly detected up to 50 keV. The {\it IBIS} sky image of \rxf, accumulated over 14 years of operations, demonstrates two extended hard X-ray sources. These sources are spatially consistent with  northwest and southwest rims of \rxf\ and are also clearly visible at energies below 10~keV with \xmm. This points to a single emission mechanism operating in soft and hard X-rays.  The \intl\ 17--120~keV spectrum of \rxf\ is characterized by a power-law continuum with the photon index of $\Gamma\approx3$, that is significantly softer than $\Gamma\approx2$ determined by \xmm\ in the 1--10~keV energy range, suggesting a progressive steepening of the spectrum with the energy.

\end{abstract}


%%%%%%%%%%%%%%%%% BODY OF PAPER %%%%%%%%%%%%%%%%%%

\section{Introduction}

Supernova remnants (SNRs) are well-known accelerators of cosmic rays \citep[CRs, see, e.g.,][for a review]{2013A&ARv..21...70B}. \mbox{RX~J1713.7--3946} (hereafter \rx), also known as  \mbox{G347.3--0.5} \citep{1999ApJ...525..357S}, is one of the best studied young shell-type SNRs. It  was discovered in the Scorpius constellation during the \rosat\ All Sky Survey in soft X-rays \citep{1996rftu.proc..267P}. In this band, the source has a slightly elliptical shape with the maximum size of $\mysim70'$. Based on \chan\ observations, \cite{2003A&A...400..567U} revealed  bright filamentary structures in \rx. A double-shell structure was also detected in the western part of \rx\ with an enhanced absorption along this edge \citep{2004A&A...427..199C}, likely associated with nearby molecular clouds \citep{2001ApJ...562L.167B,2005A&A...431..953H, 2004A&A...427..199C, 2003PASJ...55L..61F}. The distance to \rx\ was estimated to be $\mysim1$~kpc based on the molecular gas observations \citep{2003PASJ...55L..61F, 2005ApJ...631..947M}. Such a location most likely coincides with the historical supernova SN~393, which age is about 1600~years \citep{1997A&A...318L..59W}.

\rx\ is a bright source in X-ray and gamma energy bands. Its interior regions exhibit a faint thermal component with Ne Ly$\alpha$ and Mg He$\alpha$ emission lines attributed to the reverse-shocked supernova ejecta \citep{2015ApJ...814...29K}. The X-ray emission of \rx\ is dominated by the synchrotron radiation of electrons in shell regions \citep{1997PASJ...49L...7K,1999ApJ...525..357S} accelerated up to multi-TeV energies at the supernova shock \citep{1995Natur.378..255K,2007A&A...465..695Z}.

The X-ray synchrotron spectrum in the soft 0.7--7.0~keV energy band is well described by the absorbed power-law model with the photon index $1.8<\Gamma<3.0$  and the hydrogen column density $N_{\rm H}=(0.4-1.3)\times10^{22}$~cm$^{2}$ varying within a selected SNR region \citep{2018PASJ...70...77O}. Similar results were obtained  with \xmm\ \citep{2009A&A...505..157A} and {\it Suzaku XIS} \citep{2008PASJ...60S.131T,2008ApJ...685..988T}. Hard X-rays from the  \rx\ have been detected by {\it Suzaku HXD} up to 40 keV with the spectrum well described by a power-law with the photon index of $\Gamma\sim3$, which is steeper than that measured for energies below $\mysim10$~keV \citep{2008ApJ...685..988T}.

The gamma-ray emission from SNR \rx\ was first detected with {\it CANGAROO} Cherenkov telescopes \citep{2000A&A...354L..57M,2002Natur.416..823E}. Further observations of \rx\ were carried out with {\it H.E.S.S.} gamma-telescopes \citep{2004Natur.432...75A,2006A&A...449..223A,2007A&A...464..235A}. Based on {\it H.E.S.S.} data,  \cite{2009A&A...505..157A,2018A&A...612A...6H} obtained the photon index estimate of $\mysim2$ in the 0.2--30~TeV energy band. The similarity of the SNR shell-like morphology in soft X-rays and gamma-rays suggests a common emission mechanism in both bands, presumably due to cosmic ray particles accelerated at the shocks.

Despite the large number of observatories operating in hard X-rays, there is still no detailed information about the hard X-ray morphology and spectrum of \rx\ at energies above $\mysim10$~keV. As a part of all-sky survey, \cite{2007A&A...475..775K} reported  the first detection of the \rx\ in the hard X-ray band with the {\it IBIS} coded-mask telescope on board the {\it INTEGRAL} gamma-ray observatory. It was shown that the $17-60$~keV sky image of \rx\ exhibits a clear extended structure spatially consistent with the soft X-ray morphology revealed by ROSAT in the $0.5-2.5$~keV energy band. Later, \cite{2008PASJ...60S.131T} presented the X-ray spectrum of the southwestern part of \rx\ measured with the non-imaging hard X-ray instrument {\it Suzaku HXD}. However, the full $25'\times25'$ FOV of {\it HXD} covers only a part of \rx\ and the spectral analysis requires  non-trivial modeling of the astrophysical background. Due to its large extent of $\mysim70'$, \rx\ is also a complicated target for the  hard X-ray focusing telescope \nus\ \citep{Harrison_2013}, with the $13'\times13'$ FOV, prescribed by Wolter Type 1 design. \cite{Tsuji_2019} presented the first direct X-ray image of the northwest rim of  \rx\ at energies above 10~keV obtained with \nus. However, the full \nus\ survey of the entire remnant is strongly limited by ghost-ray contamination and the known stray-light issues \citep{2017JATIS...3d4003M}.

Given the above difficulties of observing \rx\ in the hard X-ray band, \intl\ provides a reasonable trade-off between the size of the source ($\mysim70'$) and the angular resolution of the {\it IBIS} telescope ($12'$ FWHM). 

In this paper, we present the first detailed spatial and spectral study of \rx\ with the {\it INTEGRAL/IBIS}, using the significantly increased exposure time on the source since the work by \cite{2007A&A...475..775K}. The paper is structured as follows: in Sect.~\ref{sect:obs}, we describe the observations and data reduction; the \rx\ morphology and spatial analysis of the data in soft and hard X-rays are presented in Sect.~\ref{sect:morph} and \ref{sect:2d}, respectively; Sect.~\ref{sect:spec} contains the procedure of the \rx\ spectral analysis; the obtained results are discussed and summarized in Sections~\ref{sect:discussion} and \ref{sect:summary}, correspondingly.

\section{Observations and data analysis}
\label{sect:obs}

This work is based on data acquired with the \ib\ coded-mask telescope \citep{2003A&A...411L.131U} on board the \intl\ gamma-ray observatory \citep{2003A&A...411L...1W} from December 2002 to March 2017. We followed the data analysis procedure described in \cite{2014Natur.512..406C,2017MNRAS.470..512K} to produce sky images of \rx\ in 17--27, 27--36, 36--50, 50--120 and 17--60~keV energy bands.

In order to extend the image and spectral analysis of \rx\ to the standard X-ray energy band of $1-10$~keV, we use all available \xmm\ archival observations of \rx\ from  2001 to 2017 with the total (effective) exposure of 711~ks (549~ks) as listed in Table~\ref{tab:xmm_obs}. In this work, we use data from {\it EPIC/MOS} cameras only. We followed the analysis procedure described in \cite{2003ApJ...590..225C} to obtain a background-subtracted, exposure and vignetting-corrected image of the entire region of SNR \rx\ in the 1--10~keV energy band.

\begin{table}
\noindent
\centering
\caption{The list of the \xmm\ observations used in this work.}
\label{tab:xmm_obs}
\centering
\vspace{1mm}
\begin{tabular}{c|c|c|c}
    \hline\hline
    ObsID & Date and time & Full  & Effective  \\
    & of observation& exposure, ks & exposure, ks \\
    \hline
	0093670101 & 2001-09-05 02:27:41 & 15.0               & 3.1                    \\
	0093670201 & 2001-09-05 07:21:05 & 15.0               & 10.4                   \\
	0093670301 & 2001-09-07 23:55:07 & 16.2               & 15.4                   \\
	0093670401 & 2002-03-14 15:52:41 & 13.7               & 12.8                   \\
	0093670501 & 2001-03-02 17:39:37 & 14.8               & 13.5                   \\
	0203470401 & 2004-03-25 08:02:28 & 16.9               & 16.5                   \\
	0203470501 & 2004-03-25 13:35:49 & 16.9               & 15.2                   \\
	0207300201 & 2004-02-22 14:15:54 & 34.2               & 16.8                   \\
	0502080101 & 2007-09-15 03:41:24 & 34.9               & 19.5                   \\
	0502080201 & 2007-09-03 06:27:15 & 25.4               & 2.9                    \\
	0502080301 & 2007-10-03 04:53:43 & 25.1               & 5.7                    \\
	0551030101 & 2008-09-27 16:44:57 & 24.9               & 23.6                   \\
	0722190101 & 2013-08-24 20:48:23 & 138.9              & 128.8                  \\
	0740830201 & 2014-03-02 07:25:30 & 140.8              & 105.7                  \\
	0743030101 & 2015-03-10 21:36:40 & 83.9               & 74.3                   \\
	0804300801 & 2017-08-30 16:32:02 & 47.8               & 45.0                   \\
	0804300901 & 2017-08-29 15:10:48 & 47.0               & 39.5                   \\
    \hline
\end{tabular}\\
\vspace{3mm}
\end{table}

\section{\rx\ hard X-ray morphology}
\label{sect:morph}

The \textit{IBIS} telescope is a soft gamma-ray (20~keV -- 10~MeV) instrument constructed in coded-mask design with the low-energy detector layer \textit{ISGRI} \citep{2003A&A...411L.141L} and high-energy layer {\it PICSIT} \citep{2003A&A...411L.189D}. The coded-aperture imaging system with a tungsten mask located at 3.2~m above the detector plane provides the angular resolution of $12'$. Since the coded-mask  telescope design does not provide a direct imaging, deconvolution-based sky image reconstruction procedures are applied to obtain relative positions of X-ray sources in the FOV. This approach is not suitable for observing sources with an angular size greatly exceeding the telescope angular resolution. However, if the source size is slightly larger than the full width at half maximum (FWHM) of the telescope point spread function (PSF), it is possible to obtain some limited information about the morphology of the extended source emission. For instance, \cite{2007A&A...470..835E,2008A&A...479...27E,2008ApJ...687..968L} effectively utilized {\it IBIS} data to study  the extended hard-X-ray emission of  galaxy clusters.

\begin{figure}
  \centering
  \includegraphics[width=0.9\linewidth]{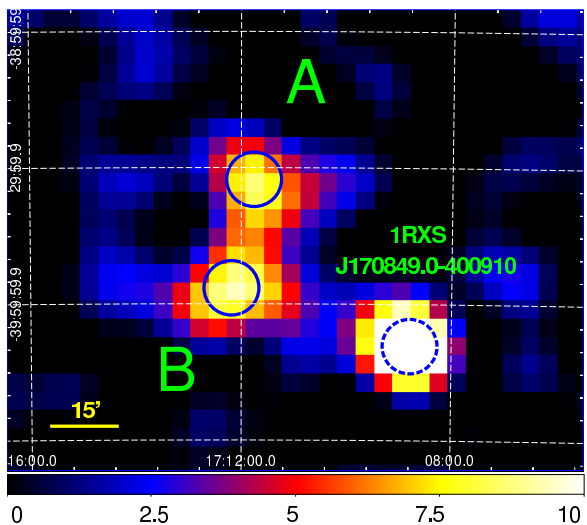}
  \caption{{\it IBIS/ISGRI} image of \rx\ in the $17-60$~keV energy band. The image is convolved with the Gaussian with $\sigma = 5'$, representing the instrumental PSF. The pixel image is constructed in terms of $S/N$ with the color map showing values from 0 to 10. Two hard X-ray excesses of SNR are marked as A and B, and correspond to the northwest and the southwest rims of \rx, respectively. Blue solid circles with $6'$ radius are placed at corresponding centroid coordinates (Sect.~\ref{sect:morph}). Blue dashed circle with the same radius shows the position of the X-ray source \rxs.}
  \label{fig:im_all}
\end{figure}

The {\it IBIS/ISGRI} $17-60$~keV image of \rx\ is shown in Fig.~\ref{fig:im_all}. The surface brightness of \rx\ is dominated by two hard X-ray excesses referred later as A and B, which have been detected at the significance level of 8.7$\sigma$ and 9.3$\sigma$, respectively, which is a factor of $\mysim2$ higher than the detection significance reached in \cite{2007A&A...475..775K}. The statistical significance of the total extended emission is $>18\sigma$. The improvement in sensitivity is consistent with the increased {\it IBIS} exposure for this region from 5.3~Ms \citep{2007A&A...475..775K} to $\mysim24$~Ms \citep{2017MNRAS.470..512K} over the period of 10 years. The centroid position of A and B excesses have been determined, respectively, at RA=$17^{\rm h}11^{\rm m}45\fs6$, Dec.=$-39\degr32'35\farcs0$ and RA=$17^{\rm h}12^{\rm m}11\fs2$, Dec.=$-39\degr56'27\farcs4$. The localization error of point X-ray sources detected with {\it IBIS/ISGRI} depends on the source significance \citep{2003A&A...411L.179G}. Given the detection significance for A and B at the level of $\mysim10\sigma$ each, the corresponding 68\% confidence interval of centroid position is $1'.5$ \citep{2007A&A...475..775K}, provided that the extent of A and B is not very large.

\begin{figure}
  \centering
  \includegraphics[width=0.9\columnwidth]{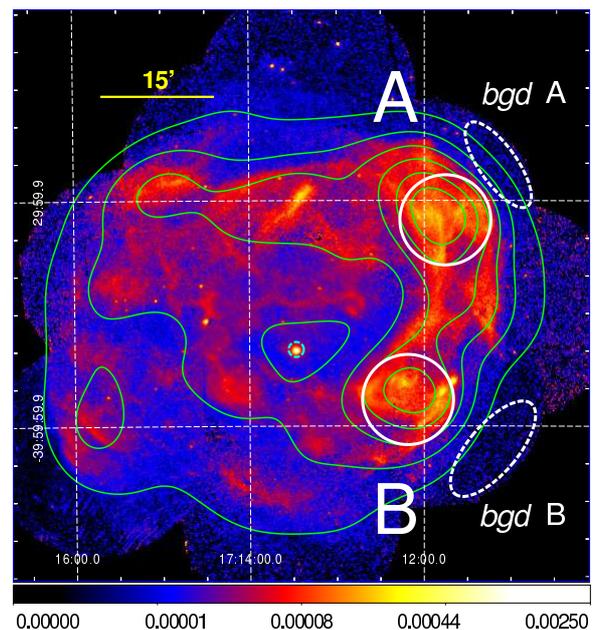}  
  \caption{The background-subtracted, exposure and vignetting-corrected \xmm\ (\mos) surface brightness maps of \rx\ in the 1--10~keV energy range. The units of the image are counts~s$^{-1}$cm$^{-2}$arcmin$^{-2}$. The image was smoothed with the Gaussian kernel with $\sigma=2''$. White solid circles mark the positions of {\it IBIS/ISGRI} hard excesses A and B. Dashed elliptical sky regions show regions used for the background evaluation in the spectral analysis. Green  contours denote isophotes of the same original \xmm\ image convolved with the {\it IBIS} instrumental PSF (the 2D Gaussian with $\sigma=5'$), after removal of the brightest point-like source \wga\ shown by the dashed circle.}
  \label{fig:im_xmm}
\end{figure}

\begin{figure*}
  \centering  
  \includegraphics[width=0.98\linewidth]{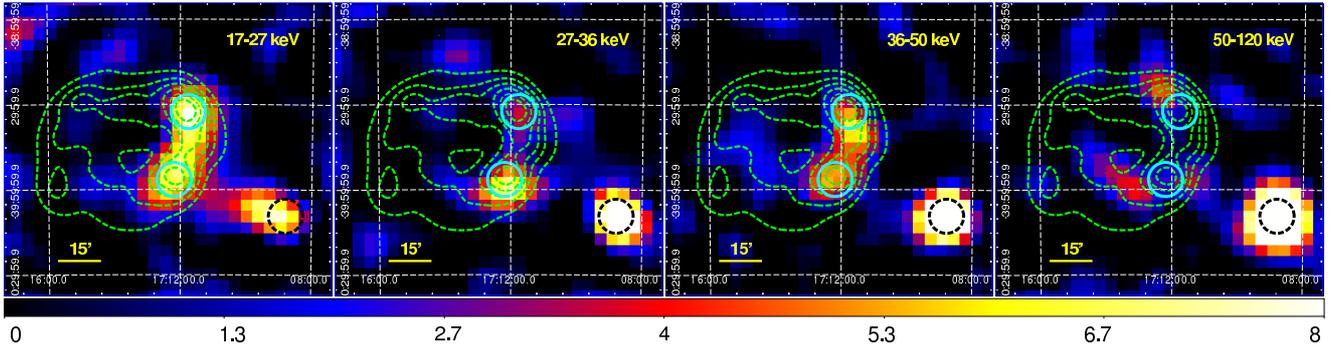}
  \caption{{\it IBIS/ISGRI} images of \rx\ in different energy bands. The images are in units of signal-to-noise ratio. Dashed green contours represent the \xmm\ $1-10$~keV surface brightness distribution convolved with the \ib\ PSF. The position of the X-ray source \rxs\ is marked as a black dashed circle.}
  \label{fig:im_4b}
\end{figure*}

Fig.~\ref{fig:im_xmm} shows the detailed $1-10$~keV \xmm\ image of \rx\ with positions of hard X-ray sources A and B. As seen from the image, the positions of A and B are spatially consistent with the brightest parts of the northwest and the southwest rim of the SNR, respectively. Fig.~\ref{fig:im_4b} shows the {\it IBIS/ISGRI} image of \rx\ in the 17--27, 27--36 and 36--50~keV energy bands. The shell-like structure of the SNR is readily visible in the softest {\it IBIS/ISGRI} $17-27$~keV band. It is also clearly seen in hard X-ray band, up to $\mysim50$~keV. Note that source A appears fainter than B in the 27--36~keV energy bands. This could  simply be the result of statistical fluctuations or an artifact of the image reconstruction method for a weak source in a densely populated environment of nearby sources \citep[see, e.g.][]{2010A&A...519A.107K}, which might increase the noise level. At higher energies of 50--120~keV, the image is dominated by noise, although a hint of a positive excess over the \rx\ shell region might be present, albeit with low significance. 

The {\it IBIS/ISGRI} hard X-ray image of \rx\ also contains a bright X-ray source with the flux\footnote{The flux units are in mCrab, where 1~mCrab in the 17--60~keV energy band corresponds to the flux of $1.43\times10^{-11}$~erg~s$^{-1}$~cm$^{-2}$ for a source with a spectrum similar to that of the Crab Nebula, represented as $10.0 \times E^{-2.1}$~keV~photons~cm$^{-2}$~s$^{-1}$~keV$^{-1}$.} of $1.10\pm0.05$~mCrab ($17-60$~keV) located $\mysim41'$ from source B in the south-west direction. Its centroid position RA=$17^{\rm h}08^{\rm m}47\fs5$, Dec.=$-40\degr09'14\farcs4$ is consistent with the \rxs\ source discovered in the \rosat\ All-Sky Survey \citep{1999A&A...349..389V} and identified as an anomalous X-ray pulsar using the {\it ASCA} observations \citep{1997PASJ...49L..25S}.

\subsection{2D image analysis}
\label{sect:2d}

\begin{figure}
  \centering  
  \includegraphics[width=0.9\columnwidth]{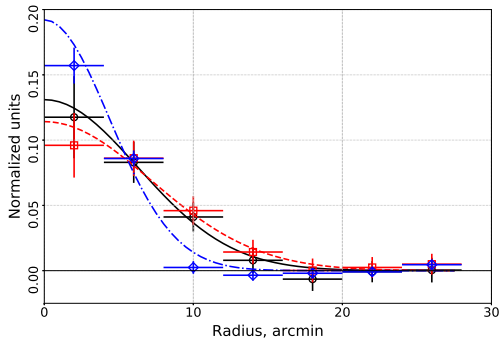}
  \caption{Radial profiles of the sources A (circles), B (squares), and \rxs\ (diamonds). Solid, dashed and dash-dotted lines show the corresponding best-fit Gaussian functions with standard deviations of $\sigma=$ $6'.3^{+1'.3}_{-1'.1}$, $7'.2^{+1'.5}_{-1'.2}$ and $4'.4\pm0'.3$ for the sources A, B, and \rxs, respectively.}
  \label{fig:prof_src}
\end{figure}

In the standard X-ray band, the brightest regions A and B are clearly extended, with a characteristic size of $\mysim12'$, i.e. comparable to the PSF FWHM of the IBIS telescope. Assuming that the same emission mechanism is operating in the hard band, we expect their extended nature to show up in the IBIS data too. To verify this conjecture, we compare 17--60~keV radial profiles of the sources A and B with that of the point-like source \rxs. The latter is expected to be consistent with the PSF of the {\it IBIS} telescope. The radial profiles were extracted relative to the corresponding centroid position, and re-normalized to the unit integral value for convenience.  The extracted radial profiles are shown in Fig.~\ref{fig:prof_src}. We fitted the profiles with the Gaussian function with a center fixed at the zero value. As a result, we obtained the following Gaussian standard deviations $\sigma_{\rm A}=6'.3^{+1'.3}_{-1'.1}$, $\sigma_{\rm B}=7'.2^{+1'.5}_{-1'.2}$ and $\sigma_{\rm 1RXS}=4'.4\pm0'.3$, respectively, for A, B, and \rxs. The width of the latter is consistent with $\sigma=5'$ of the {\it IBIS} PSF. The spatial extent of A and B is systematically (albeit, marginally) larger than the {\it IBIS} PSF, suggesting that the {\it IBIS/ISGRI} image of \rx\ follows the morphology of the SNR, and it is not consistent with two hard X-ray point-like sources seen in projection on \rx\ shell.

\begin{figure}
\centering
\center{\includegraphics[width=0.75\linewidth]{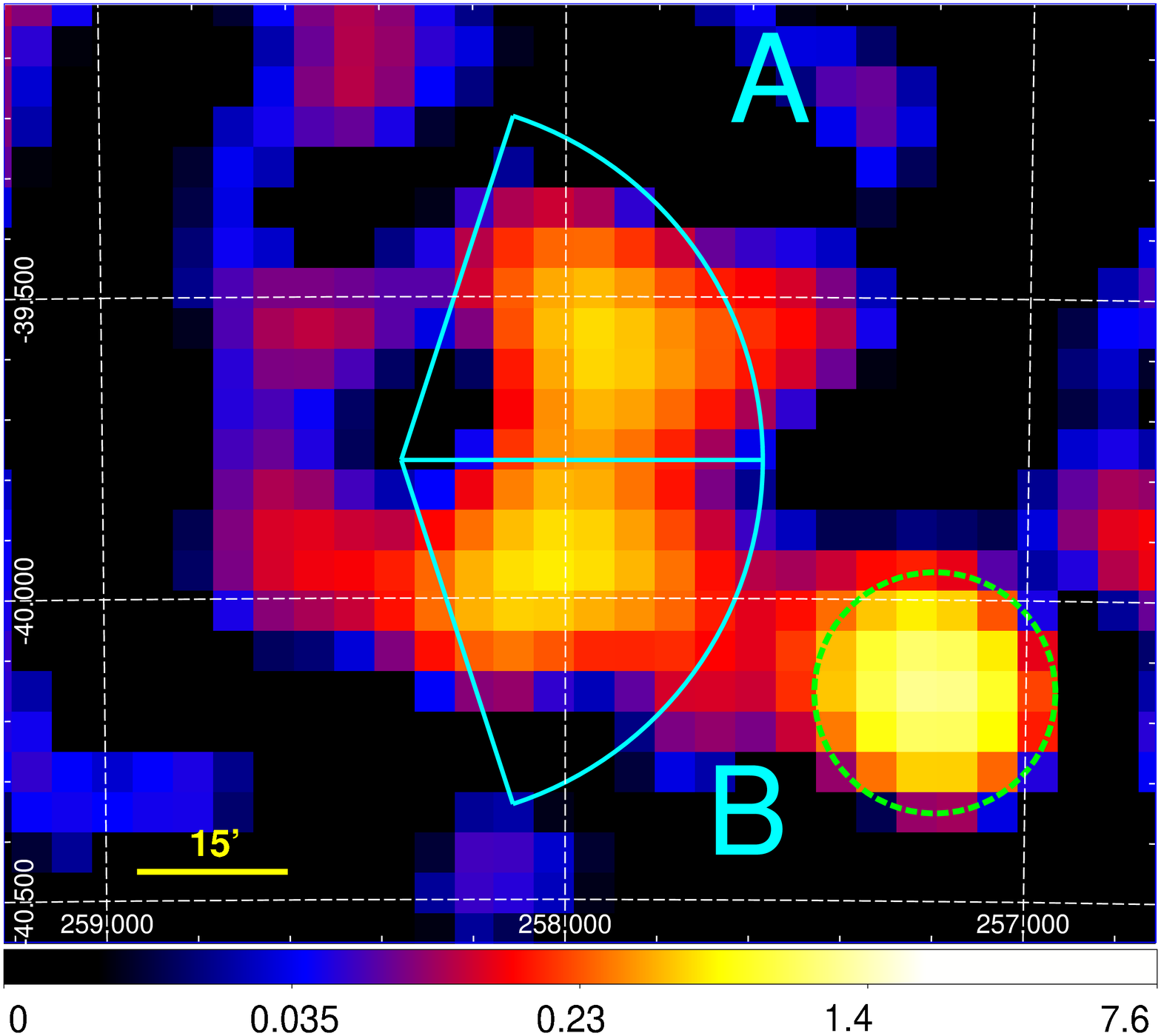}}
\center{\includegraphics[width=0.75\linewidth]{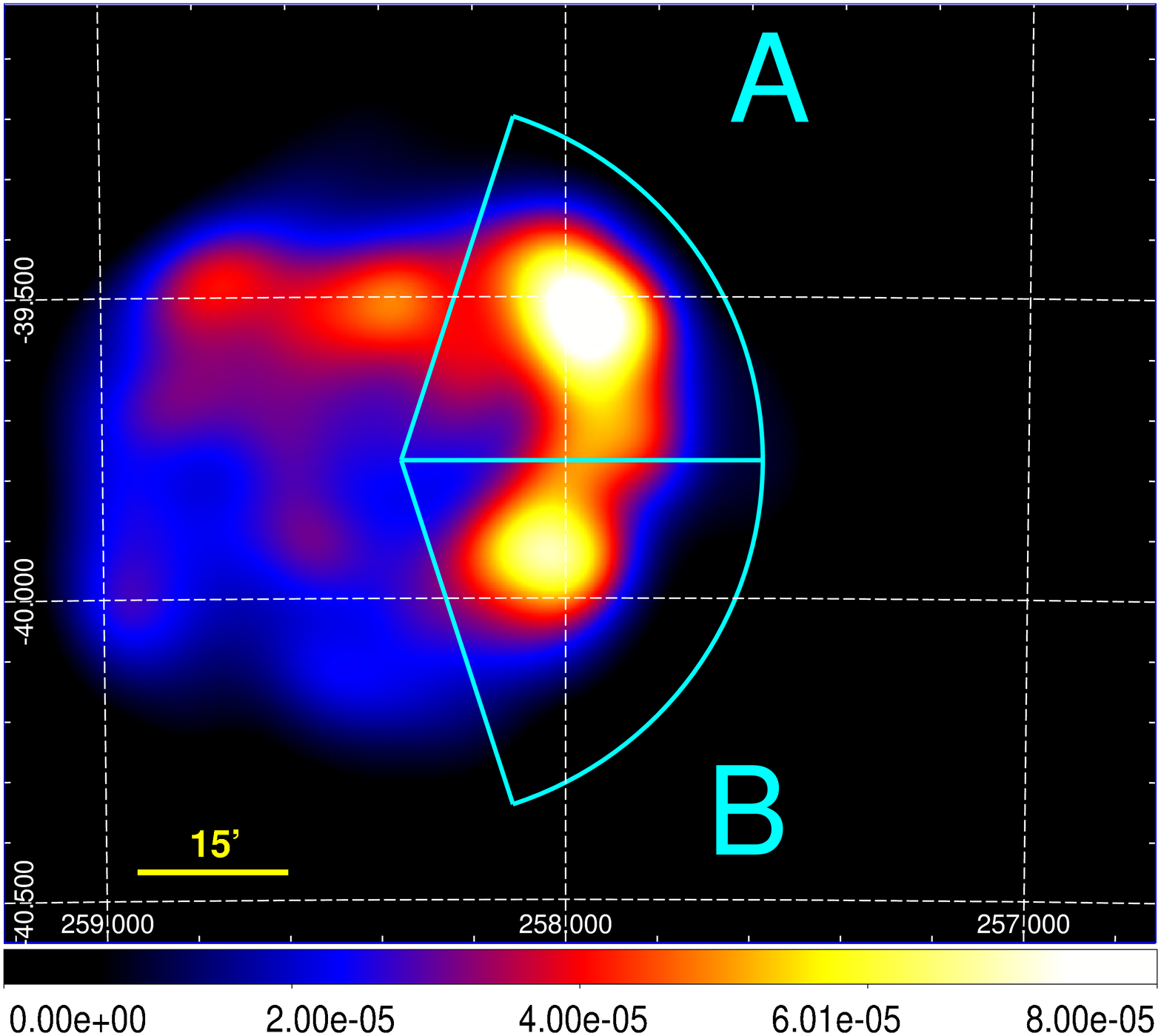}}
\caption{Top: the 17-60~keV \ib\ image of \rx\ in mCrab convolved with the PSF. The green dashed circle with the radius of $R=12'$ shows the \rxs\ region excluded from the spatial analysis. Bottom: the 1--10~keV \xmm\ \mos\ surface brightness image of \rx\ convolved with the \ib\ PSF. In the both images, sectors, for which radial profiles were constructed, are highlighted with blue color.}
\label{fig:sect}
\end{figure}

However, since \rx\ has a shell-like typical SNR morphology, we then extracted the {\it IBIS/ISGRI} radial profiles for A and B sources relative to the center of the shell-like structure located at RA=$17^{\rm h}13^{\rm m}25\fs2$ and Dec.=$-39\degr46'15\farcs6$. The center coordinates were taken from a three-dimensional spherical shell model evaluated on {\it H.E.S.S.} atmospheric Cherenkov telescope data \citep{2018A&A...612A...6H}. The radial profiles were extracted  from both the {\it IBIS/ISGRI} and \xmm\ (\mos) images in two corresponding sectors with the opening angle of $72^{\circ}$ as shown in Fig.~\ref{fig:sect}. To compare \ib\ and \xmm\ radial profiles, we renormalized profiles in range $14'<R<40'$ (sector A) and $10'<R<40'$ (sector B). Note that the \xmm\ image was convolved with the {\it IBIS} PSF. Before the convolving procedure, we also removed from the \xmm\ image a circular region $R=0'.9$ with the bright X-ray source \wga, which is probably a neutron star and the central compact object of SNR \rx\ \citep{2003ApJ...593L..27L,2004A&A...427..199C}.

\begin{figure}
\centering
\includegraphics[width=0.85\linewidth]{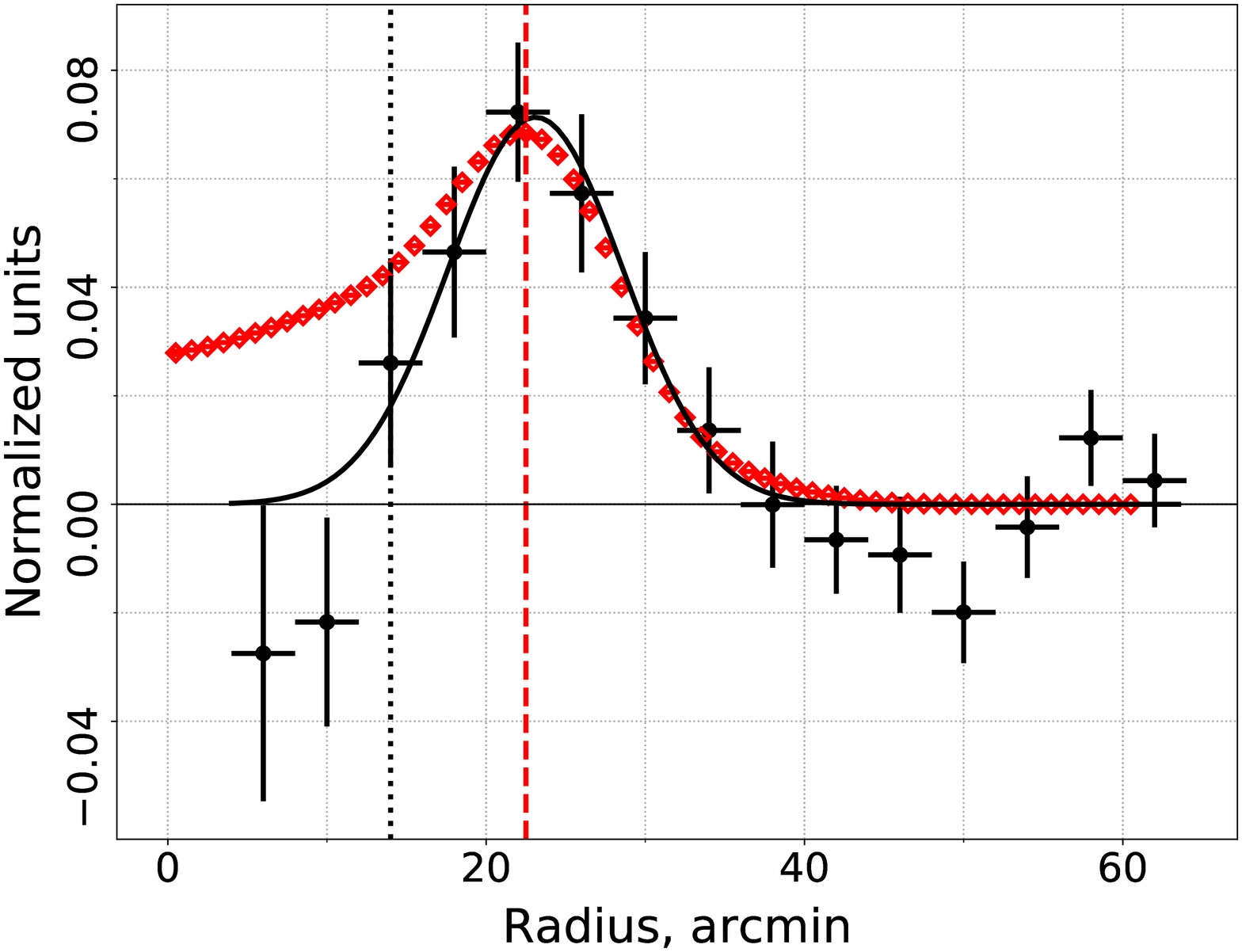}
\includegraphics[width=0.85\linewidth]{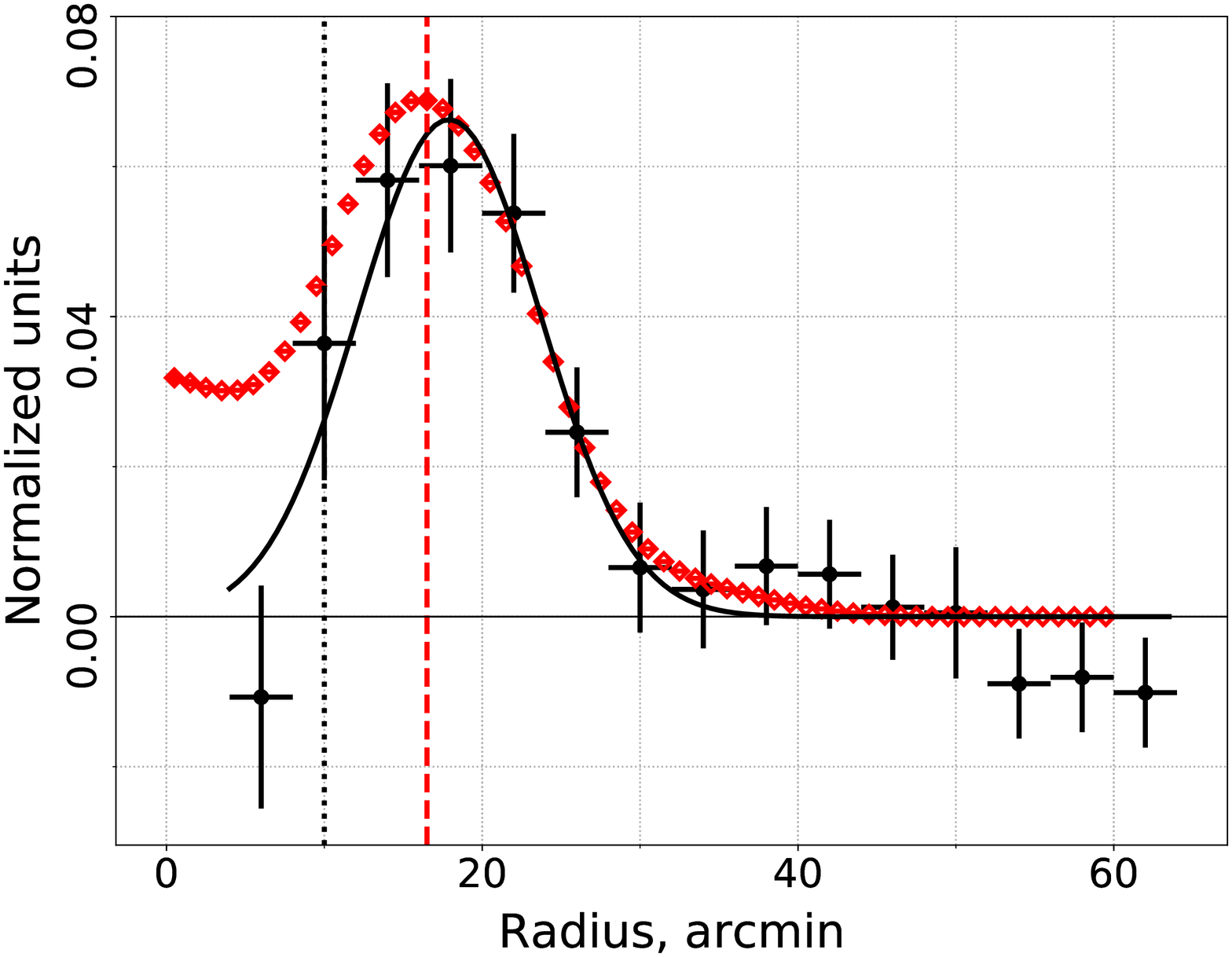}
\caption{\intl\ (circles) and \xmm\ \mos\ (squares) radial profiles of sectors A (top panel) and B (bottom panel). In each panel, the solid line represents the best-fit Gaussian model for the \intl\ data. The positions of the \xmm\ emission peaks are marked as vertical dashed lines. The radial profiles have been re-normalized to the unit area starting from the radius marked by the vertical dotted lines.}
\label{fig:prof}
\end{figure}

The extracted radial profiles of A and B are shown in Fig.~\ref{fig:prof}. We fitted the \ib\ radial profiles with a simple Gaussian function to estimate the offset of the shell-like structure relative to the center. The best-fit model parameters are listed in Table~\ref{tab:gau}. The obtained best-fit Gaussian functions for the sectors A and B have comparable widths but they are significantly shifted relative to each other, however, consistent with the corresponding peaks of the \xmm\ radial profiles.

\begin{table}
\noindent
\centering
\caption{The best-fit Gaussian model parameters for {\it IBIS/ISGRI} radial profiles within sectors A and B (Fig.~\ref{fig:sect}), and the corresponding \xmm\ peak emission offset.}
\label{tab:gau}
\vspace{1mm}
\begin{tabular}{c|c|c}
\hline
Parameter                          & Sector A        & Sector B       \\
\hline \hline
Amplitude, mCrab                   & $0.43\pm0.06$   & $0.6\pm0.08$   \\
Offset position                    & $23'.1\pm1'.1$  & $17'.9\pm1'.0$ \\
$\sigma$                           & $5'.5\pm1'.0$   & $5'.8\pm0'.9$  \\
$\chi^2_{\rm r}$/d.o.f.            & $0.95/12$       & $0.71/12$     \\
\hline
\xmm\ peak emission offset         &     $22'.5$     & $16'.5$     \\
\hline
\end{tabular}
\vspace{3mm}
\end{table}

Note that  a double-shell structure along the western limb of  the  remnant is clearly seen in the soft X-ray \xmm\ image of \rx\ (see Fig.~\ref{fig:im_xmm}). Presumably, the outer shell is caused by a forward shock wave, while the inner one is due to a shock wave reflected from the nearest molecular cloud \citep{Tsuji_2019}. Because of insufficient angular resolution ($12'$ FWHM) of \ib, we cannot directly detect a double-shell structure. However, after convolving the \xmm\ image with the \ib\ PSF, we see the similar shell-like structure dominated by two bright clumps whose positions are consistent with the \ib\ A and B hard X-ray excesses (see Fig.~\ref{fig:sect}). This agreement indicates that a double-shell structure is likely to be remaining in hard X-rays, too.

\section{Spectral analysis}   
\label{sect:spec}

X-ray spectra of supernova remnants carry an unique information about a cosmic rays acceleration at expanding shocks. In particular, the hard X-ray spectrum of \rx\  can shed light on the diffusion regime of electrons in the SNR shock region \cite{2007A&A...465..695Z}. Since the observations in the hard X-ray energy band are complicated due to many different factors (e.g., a modest angular resolution or even non-imaging, background-dominated measurements, systematic noise, stray-light, low signal-to-noise ratio, etc.), the information about hard X-ray emission of \rx\ is limited. Thanks to the long-term \intl\ observations of the Galactic Center, we now can significantly detect the \rx\ emission up to $\mysim50$~keV and improve its spectral information in hard X-rays.

\begin{figure}
  \centering  
  \includegraphics[width=0.99\columnwidth]{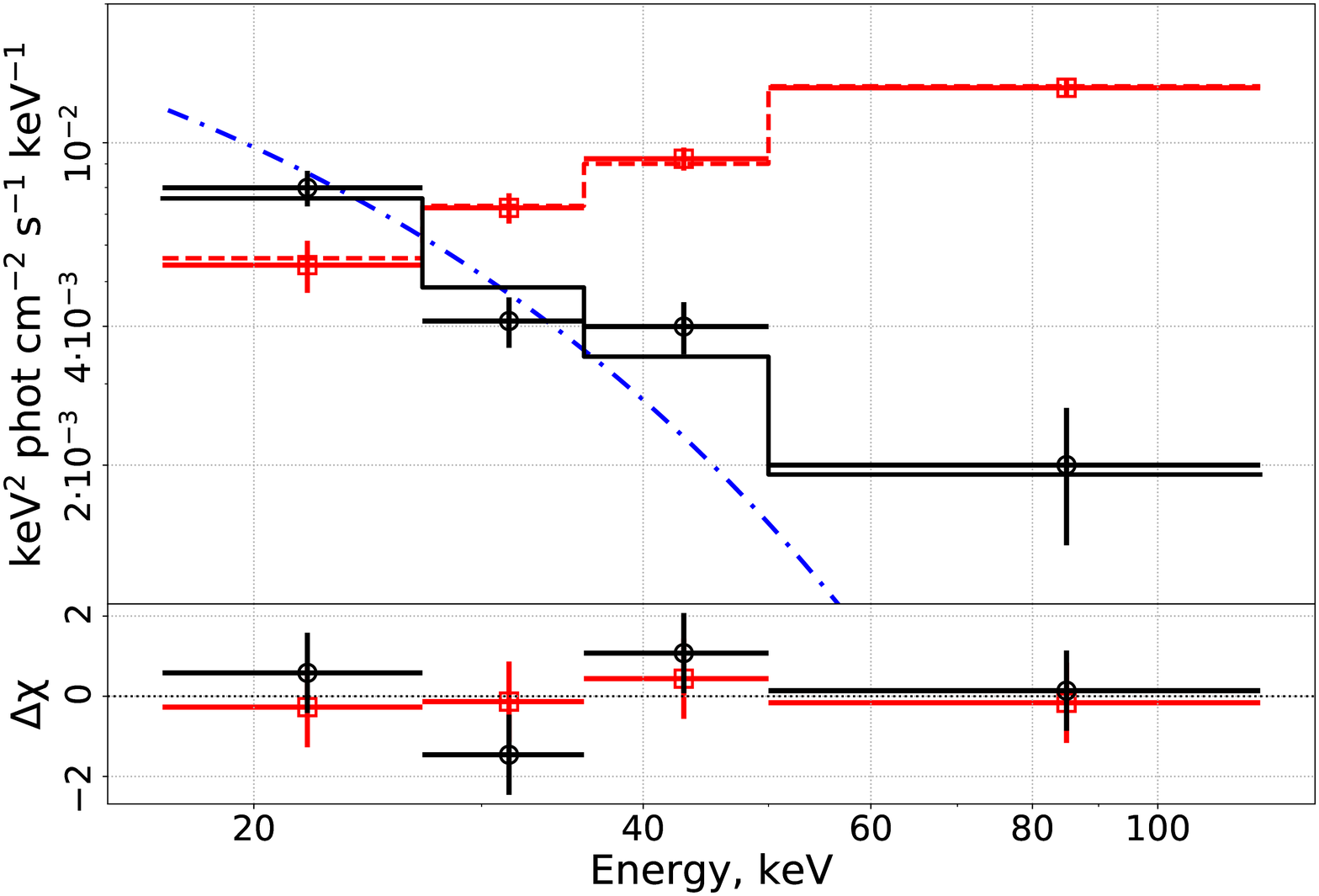}
  \caption{The summed 17--120~keV \ib\ spectrum of sources A and B is shown with black circles. The solid black line shows the power-law model, while the dash-dotted blue line shows the absorbed cutoff power-law model with the following parameters: $N_{\rm H}=1.26~10^{22}$~cm$^2$, $\Gamma=2.32$ calculated from the \xmm\ \mos\ data and fixed $E_{\rm cut}=17$~keV. The 17--120~keV \ib\ \rxs\ spectrum and its power-law model are marked with red squares and the dashed red line, respectively.}
  \label{fig:spec_ib}
\end{figure}

We extracted the \rx\ spectrum in the $17-120$~keV energy band from the {\it IBIS/ISGRI} images in fractional units of the Crab Nebula flux (mCrab, see note in Sect.~\ref{sect:morph}) which can be directly converted to  physical units. According to the matched-filter approach \citep{1995ApJ...451..542V}, the optimal way to collect the flux of a point-like source (and hence, to obtain the highest signal-to-noise ratio), is to use images convolved with the telescope instrumental PSF. For an extended source, the convolution kernel should reflect the size and morphology of the object, as it was done by \cite{2008ApJ...687..968L} to study the hard X-ray emission of the Coma cluster using {\it IBIS/ISGRI} data. For this reason, we extended the size of the Gaussian kernel from $\sigma=5'$ to $\sigma=6'.75$, which is the average value between $\sigma_{\rm A}\sim6'.3$ and $\sigma_{\rm B}\sim7'.2$ determined in Sect.~\ref{sect:2d}. To build the spectrum, we extracted fluxes of the A and B sources from the images in the corresponding energy bands. The source flux has been extracted from the positions of A and B estimated in the $17-60$~keV band, since the image in this band has the highest significance. Note that the highest $50-120$~keV energy band is dominated by the background (Fig.~\ref{fig:im_4b}). However, some hint of the positive overall excess over the entire SNR region is noticed. Finally, to make the overall spectrum of \rx\ (see Fig.~\ref{fig:spec_ib}), we summed up A and B spectra to reduce statistical errors. The resulting spectrum was fitted with a pegged power-law model in the  $17-120$~keV energy band  with the following best-fitting model parameters: $\Gamma=3.13^{+0.36}_{-0.33}$, flux $F_{\rm 17-120\ keV}=(12.0\pm1.4)\times10^{-12}$~erg~cm$^{-2}$s$^{-1}$, and fit statistics $\chi^2_{\rm r}=1.82$ (reduced $\chi^2$ for 2 d.o.f.).

To validate the obtained {\it IBIS/ISGRI} spectrum of \rx, we extract the spectrum of the nearby point source \rxs\ by the same method, except that we used the Gaussian kernel with $\sigma=5'$ as the standard \ib\ instrumental PSF. The spectrum of \rxs\ is over-plotted in Fig.~\ref{fig:spec_ib} without any re-normalization. The spectrum is well described by a simple power-law model ($\Gamma=1.3\pm0.1$, normalization at 1~keV $N_{\rm @keV}=7^{+4}_{-3}\times10^{-4}$~ph~keV$^{-1}$~cm$^{-2}$~s$^{-1}$, $\chi^2_{\rm r}=0.16$ for 2 d.o.f.), which is consistent with spectral characteristics previously obtained from the \intl\ data by \cite{2008A&A...489..263D}, indicating that our {\it IBIS/ISGRI} flux determination procedure is correct.

\subsection{IROS procedure} 

\begin{figure}
  \centering 
  \includegraphics[width=0.99\columnwidth]{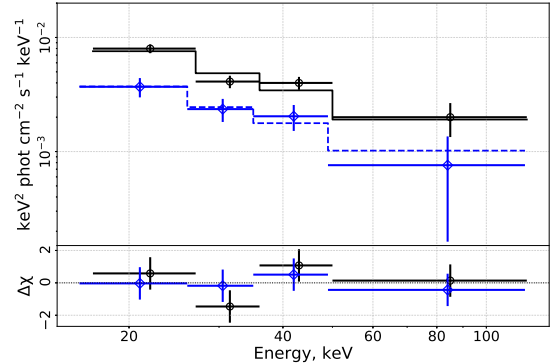}
  \caption{The summed 17--120~keV \ib\ spectrum of sources A and B and its best-fit power-law model are plotted with black circles and the solid black line, respectively. Blue diamonds and the dashed blue line are the data points and the model, respectively, calculated for the summed spectrum obtained without the IROS procedure. For convenience, this spectrum are shifted at 1 keV to the left.}
  \label{fig:spec_iros}
\end{figure}

In the \ib\ sky image reconstruction algorithm, we use the procedure of Iterative Removal Of Sources (IROS, for details see \cite{2003A&A...411L.223G,2005ApJ...625...89K,2010A&A...519A.107K}), which allows us to remove the false appearance (``ghosts'') of known X-ray sources, a by-product of a replicated \ib\ mask pattern. The IROS removes the contribution from the brightest sources first, gradually moving to the weaker ones, since the source flux is evaluated assuming that there is only one source in the \ib\ field of view.  Because of large uncertainties of the flux estimates in the individual observations for  weak sources, a systematic error appears in the final image. We extracted the combined spectrum of \rx\ without applying the IROS, to check for possible systematic effects on the final spectrum. Note that IROS was applied to all other sources in the \ib\ FOV, including nearby \rxs. The spectrum of \rx\ without IROS (shown in Fig.~\ref{fig:spec_iros}) is well described by a power-law model with the following best-fit parameters: $\Gamma=3.06^{+0.68}_{-0.57}$, $F_{\rm 17-60\ keV}=6.1\pm1.4\times10^{-12}$~erg~cm$^{-2}$s$^{-1}$, and fit statistics $\chi^2_{\rm r}=0.24$ for 2 d.o.f. Note that the photon index $\Gamma$ does not change significantly. However, the flux normalization decreases by a factor of 2. We conclude that the IROS procedure may introduce some systematic bias in the flux determination of \rx, however its impact on the spectral shape seems non-significant.

\subsection{Broad-band spectrum} 
\label{subsec:broad}

\begin{figure}
  \center{\includegraphics[width=0.99\linewidth]{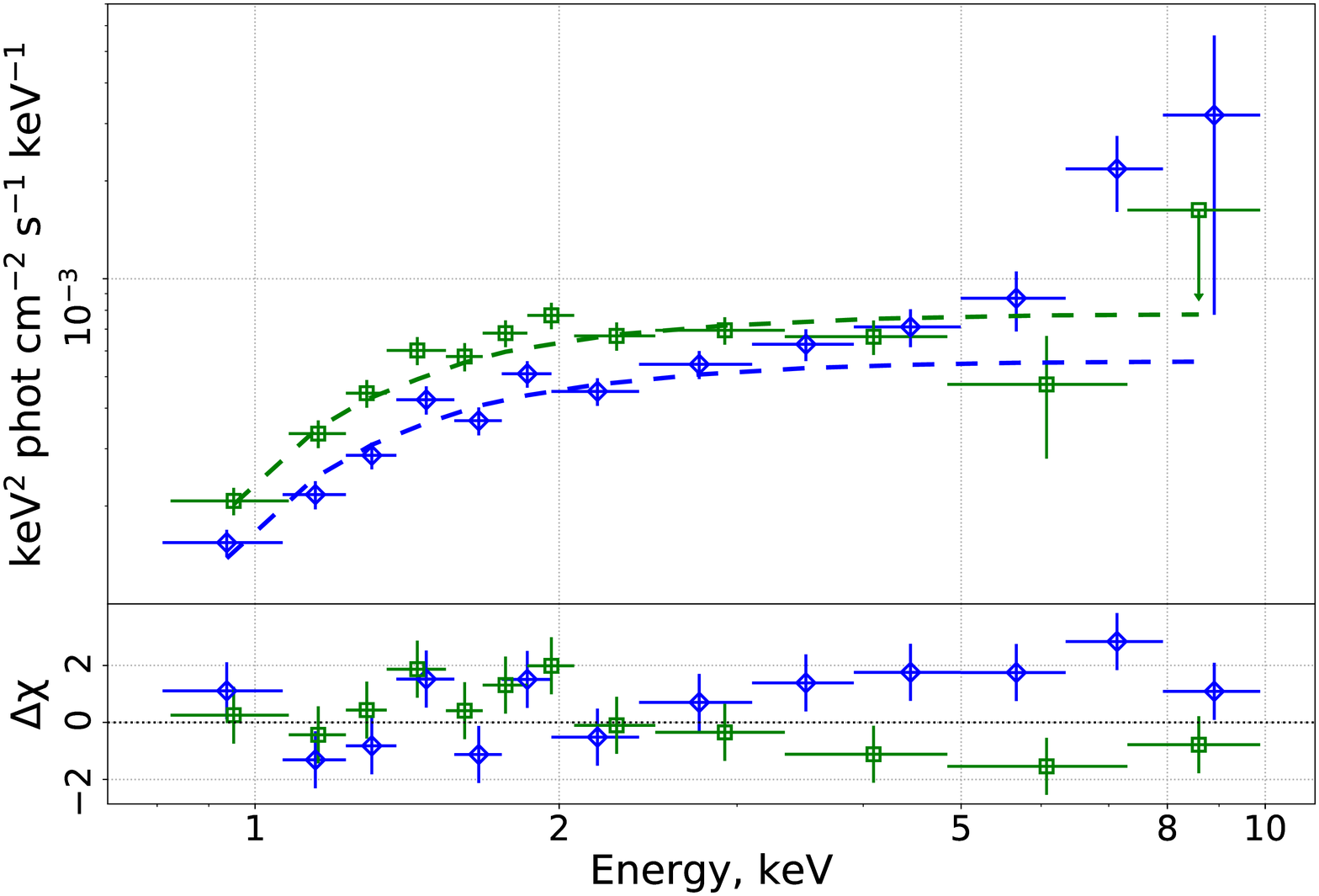}}
  \center{\includegraphics[width=0.99\linewidth]{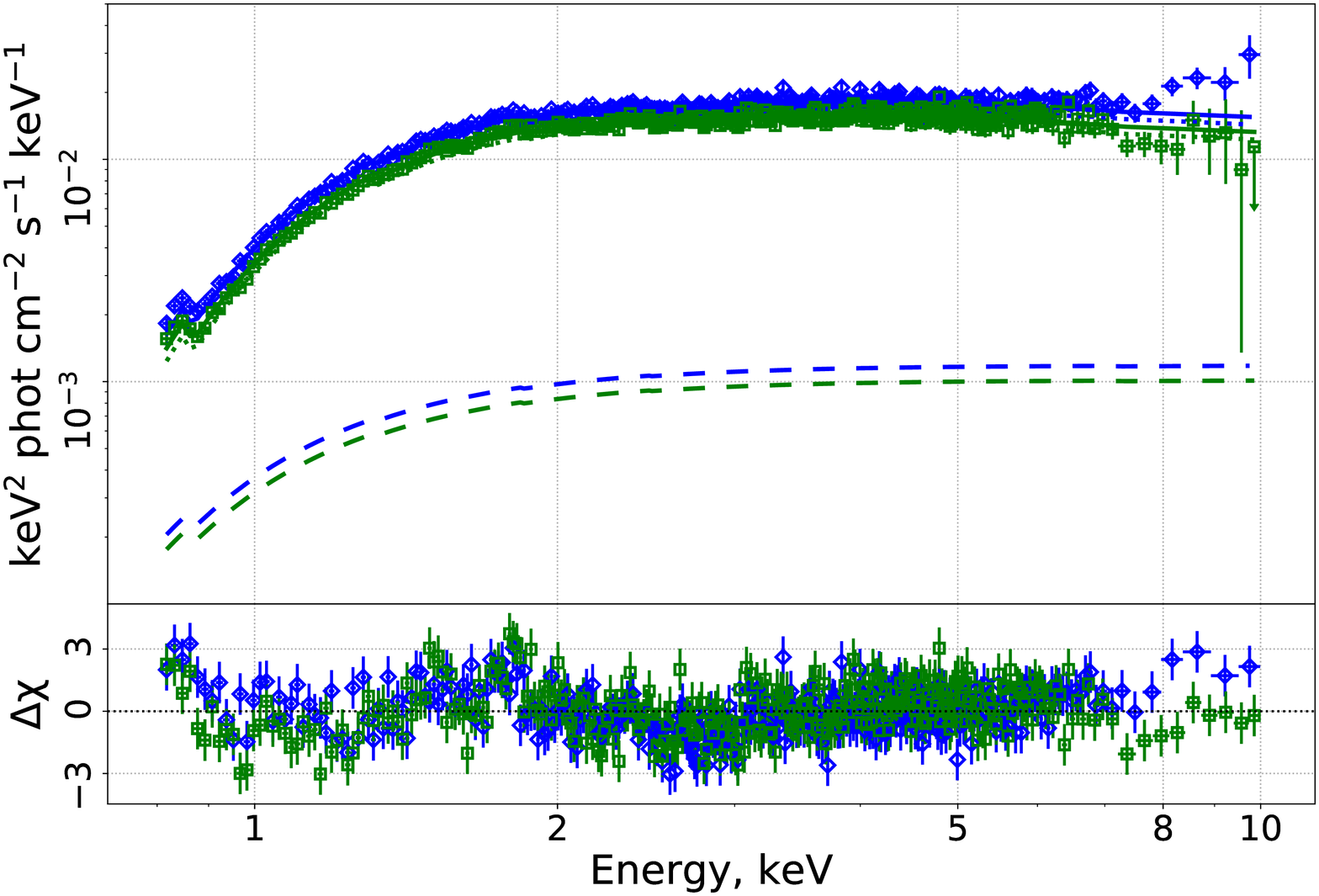}}
  \caption{Top: The 0.8--10~keV \xmm\ \mos\ background spectra extracted from the regions denoted in Fig.~\ref{fig:im_xmm} as \textit{bgd}~A (blue diamonds) and \textit{bgd}~B (green squares). Dashed lines show the model $TBabs_{bgd}*powerlaw_{bgd}$ for each spectrum. Bottom: The \xmm\ \mos\ spectra of sources A (blue diamonds) and B (green squares) in the 0.8--10~keV energy band. The source and background model components of A and B are shown by dotted and dashed lines, respectively. The corresponding total models are shown with solid lines.}
  \label{fig:spec_xmm}
\end{figure}

To obtain the broad-band X-ray spectrum of \rx, we used the \xmm\ \mos\ data in the 0.8--10~keV energy range. We chose circular regions for A and B sources (Fig.~\ref{fig:im_xmm}) with $R=6'$, what corresponds to the $12'$ FWHM of the \ib\ PSF. As background regions, we utilized two corresponding elliptical regions denoted as \textit{bgd}~A and \textit{bgd}~B in Fig.~\ref{fig:im_xmm}. The background spectra were fitted by an absorbed power-law model with the following best-fit parameters: $N_{\rm H}^{bgd} = (0.67\pm0.14)\times10^{22}$~cm$^{-2}$, $\Gamma^{bgd} = 2.0\pm0.2$,  $N^{bgd}_{\rm @keV} = (5.5\pm1.1)\times 10^{-4}$~ph~keV$^{-1}$~cm$^{-2}$~s$^{-1}$, the cross-normalization constant between \textit{bgd}~A and \textit{bgd}~B regions $C^{bgd} = 1.4\pm0.1$, and fit statistics $\chi^2_{\rm r}/d.o.f. = 0.81/330$.  Note, that for the photoelectric absorption, we used the $TBabs$ model  \citep{2000ApJ...542..914W} with the corresponding chemical abundances and absorption cross-sections by \cite{1996ApJ...465..487V}. Then, following the method described in \cite{2009A&A...505..157A}, the best-fit background model was applied in the source fitting procedure with all parameters fixed. We recalculated the normalization parameter to the areas ratio of the background \textit{bgd}~A and the source~A regions $S_{\rm A}/S_{bgd~{\rm A}}=2.15$ and fixed it at $N^{bgd}_{\rm @keV} = 1.18 \times 10^{-3}$~ph~keV$^{-1}$~cm$^{-2}$~s$^{-1}$. The final 0.8--10~keV spectra of regions A and B (Fig.~\ref{fig:spec_xmm}) are well described by an absorbed power-law model with the best-fit parameters listed in Table~\ref{tab:xmm}. 

Since we are dealing with extended sources, it is difficult to guarantee that the spectra are effectively extracted from exactly the same region, especially for the coded mask telescopes such as \ib. There are two possibilities to mitigate this problem. For instance, one can untie the normalizations of the spectra obtained with different telescopes. Another option is to fit individual spectra and then plot the slope as a function of energy. Here, we pursue the former option.

\begin{table}
\noindent
\centering
\caption{The best-fit model parameters for regions of sources A and B, measured with \xmm\ \mos\ in the 0.8--10~keV energy band. Model: $TBabs*powerlaw+TBabs_{bgd}*powerlaw_{bgd}$. Index \textit{bgd} denotes the astrophysical background model components. $C_{\rm XMM}$ is the cross-normalization constant between the source A and source B data.}
\label{tab:xmm}
\vspace{1mm}
\begin{tabular}{c|c|c}
\hline
Parameter              & Units                                        & Value          \\
\hline \hline
$N_{\rm H}^{bgd}$      & $10^{22}$~cm$^{-2}$                          & $0.67$ (fixed) \\
$\Gamma^{bgd}$         &                                              & $2.0$ (fixed)  \\
$N^{bgd,}_{\rm @keV}$  & $10^{-3}$~ph~keV$^{-1}$~cm$^{-2}$~s$^{-1}$   & $1.18$ (fixed) \\
\hline
$N_{\rm H}$            & $10^{22}$~cm$^{-2}$                          & $1.10\pm0.01$  \\
$\Gamma$               &                                              & $2.23\pm0.01$  \\
$N_{\rm @keV}$         & $10^{-3}$~ph~keV$^{-1}$~cm$^{-2}$~s$^{-1}$   & $24.4\pm0.4$   \\
$\chi^2_{\rm r}$/d.o.f.&                                              & $1.09/1141$    \\
\hline 
$C_{\rm XMM}$          &                                              & $0.858\pm0.004$\\  
\end{tabular}\\
\vspace{3mm}
\end{table}

\begin{figure}
  \centering  
  \includegraphics[width=0.99\columnwidth]{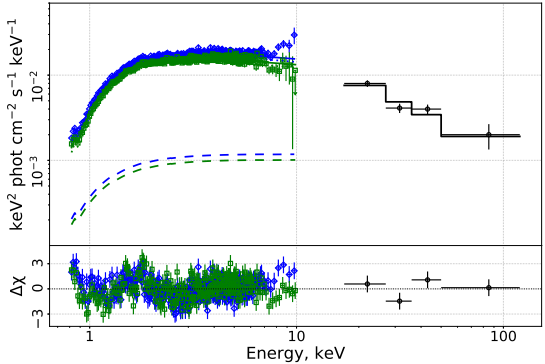}
  \caption{The broadband spectrum of sources A and B obtained by the \xmm\ \mos\ and \intl\ data at energies of 0.8--10 and 17--120~keV, respectively. The solid line shows the absorbed broken power-law model. Designations of data points and models are the same as in Fig.~\ref{fig:spec_ib} and \ref{fig:spec_xmm}.}
  \label{fig:spec_xmm_ib}
\end{figure}

The spectrum of \rx\ in the soft X-rays is well described by a power-law model with the photon index of $\Gamma\sim2$, which differs significantly from that observed in hard X-rays where $\Gamma\sim3$. This indicates a change in the slope of the power-law between somewhere around 10--20~keV. We checked whether the change in the slope is consistent with a power law with an exponential cutoff ({\it cutoffpl}) at high energies or a broken power-law ({\it bknpower}). When \xmm\ \mos\ and \ib\ data are simultaneously fitted by these models, the quality of the \ib\ data do not allow us to constrain the cutoff or break energy.  We, therefore, fixed this value to 17~keV, i.e., at the lower end of the \ib\ energy range. The best-fit model parameters are listed in  Table~\ref{tab:xmm_ibis} and the broad-band spectrum approximated by the {\it bknpower} model is shown in Fig.~\ref{fig:spec_xmm_ib}. The spectral shape of the {\it cutoffpl} model is also shown in Fig.~\ref{fig:spec_ib}. The calculated null-hypothesis probabilities of {\it cutoffpl} and {\it bknpower} models for the \ib\ data points are 0.04\% and 16\%, respectively. Thus, we come to the conclusion that the \intl\ spectrum of \rx\ in the 17--120~keV energy band favors the broken power-law model rather than the model of the power-law with an exponential cutoff (at least when the cutoff or break energy is fixed to 17~keV).

\begin{table}
\noindent
\centering
\caption{The best-fit parameters of the sources A and B spectra obtained with \xmm\ \mos\ and \intl\ in the 0.8--10 and 17--120~keV energy bands, respectively. Models: $TBabs*bknpow$ and $TBabs*cutoffpl$. $C$ is the cross-calibration constant between the \xmm\ and \ib\ data points, while $C_{\rm XMM}$ is the cross-normalization constant between the source A and source B data.}
\label{tab:xmm_ibis}
\vspace{1mm}
\begin{tabular}{c|c|c|c}
\hline
Parameter                   & Units                                        & Bknpower                                         & Cutoffpl      \\
\hline \hline
$N_{\rm H}$                 & $10^{22}$~cm$^2$                             & $1.10\pm0.01$                                    & $1.04\pm0.01$ \\
$\Gamma_1$                  &                                              & $2.23\pm0.01$                                    & $2.04\pm0.01$ \\
$E_{\rm bkn}$/$E_{\rm cut}$ & keV                                          & $17$ (fixed)                                     & $17$ (fixed)  \\
$\Gamma_2$                  &                                              & $3.13^{+0.36}_{-0.33}$                           & ---           \\
$N_{\rm @keV}$              & $10^{-3}$~ph~keV$^{-1}$~cm$^{-2}$~s$^{-1}$   & $24.4\pm0.4$                                     & $23.6\pm0.3$  \\
$C_{\rm XMM}$               &                                              & $0.858$ (fixed)                                  & $0.858$ (fixed)\\
$C$                         &                                              & $0.75^{+0.16}_{-0.14}$                           & $1.36\pm0.15$ \\
$\chi^2_{\rm r}$/d.o.f.     &                                              & $1.09/1144$                                      & $1.19/1145$   \\
\hline
\end{tabular}\\
\vspace{3mm}
\end{table}

\section{Discussion}
\label{sect:discussion}

\begin{figure*}
  \centering  
  \includegraphics[width=0.49\textwidth]{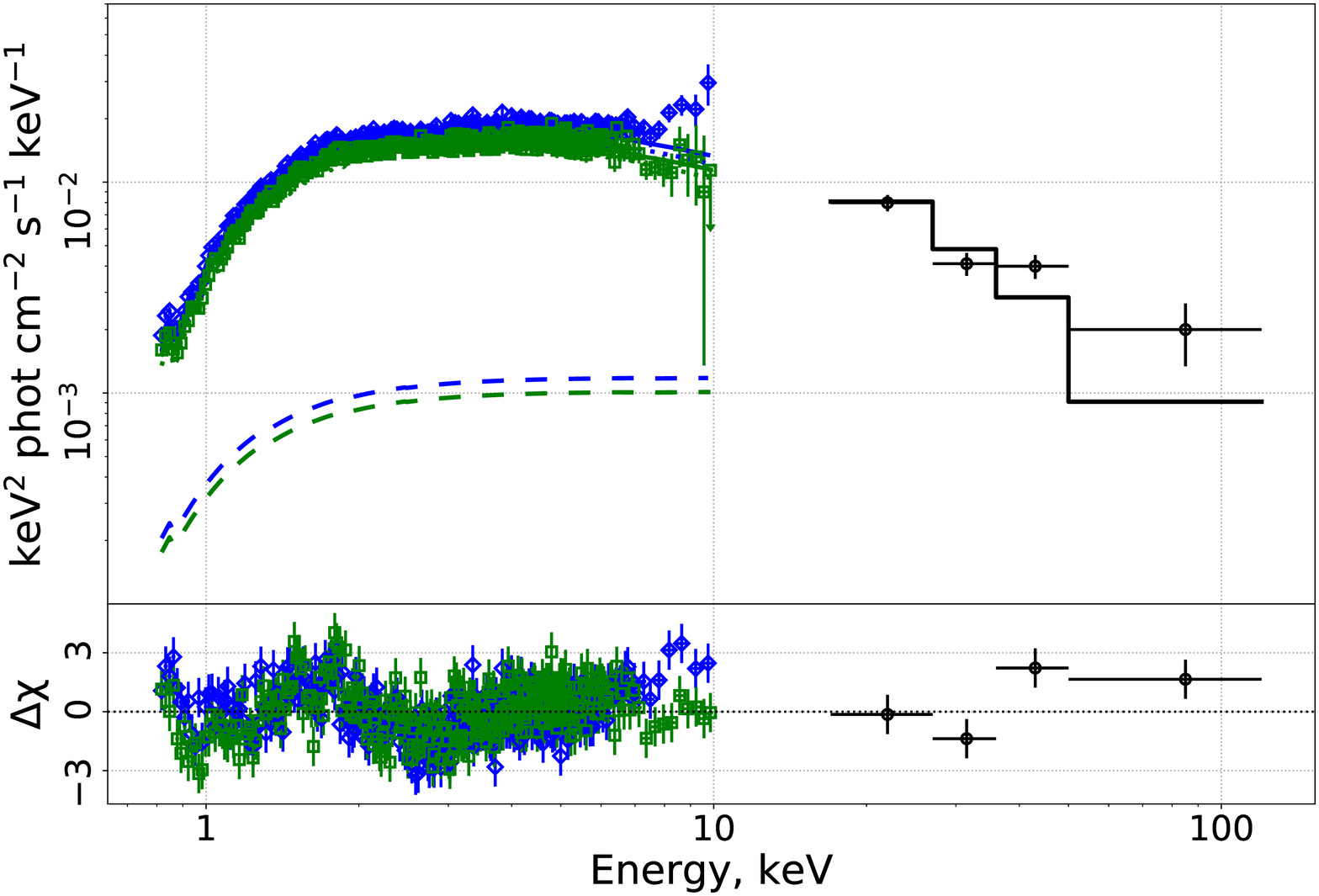}
  \includegraphics[width=0.49\textwidth]{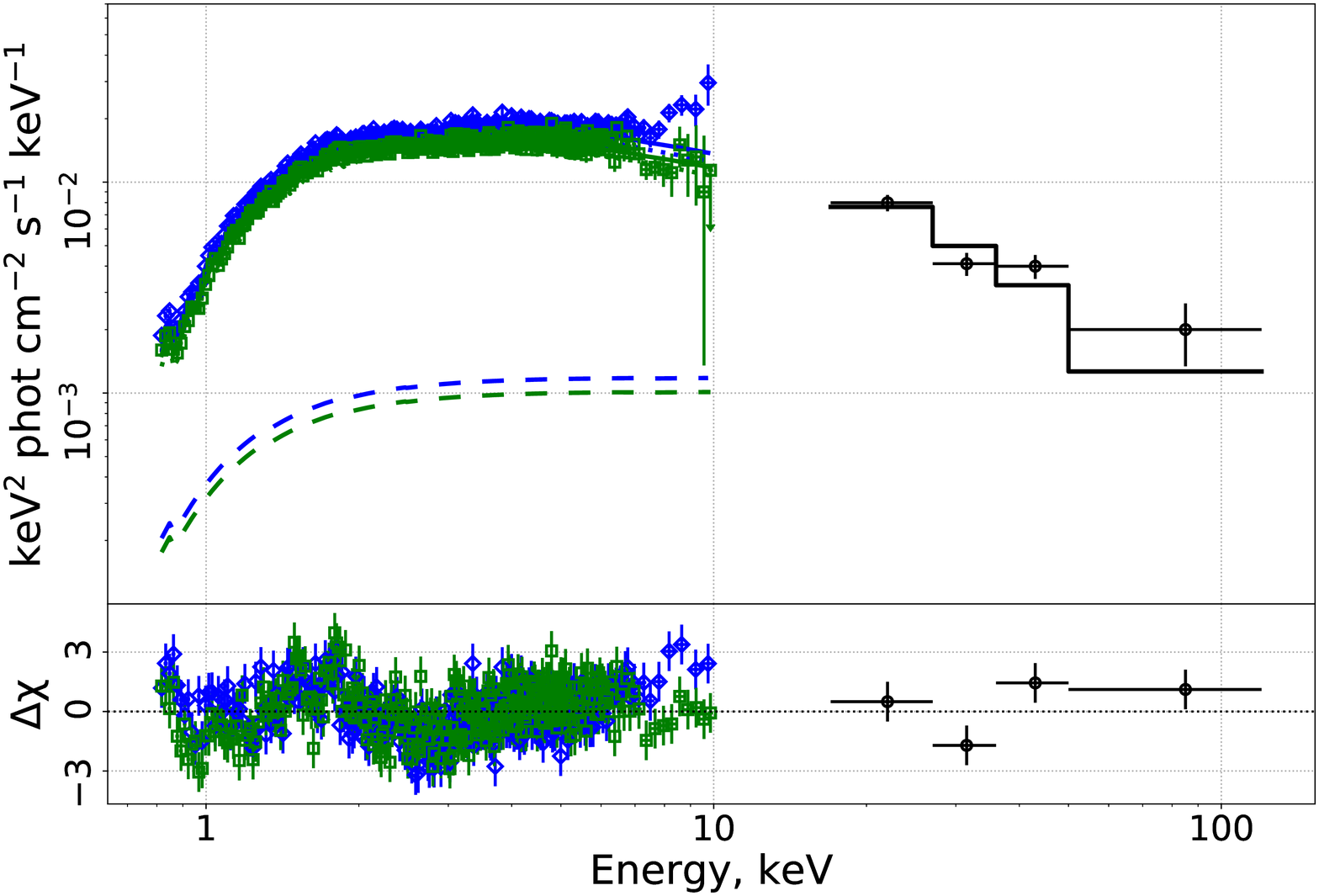}
  \caption{The broadband spectrum of sources A and B obtained by the \xmm\ \mos\ and \intl\ data at energies of 0.8--10 and 17--120~keV, respectively. The source model component is the absorbed ZA07 model for Eqs.~(\ref{eq:za1}) (left panel) and (\ref{eq:za2}) (right panel). Designations of data points and models are the same as in Fig.~\ref{fig:spec_ib} and \ref{fig:spec_xmm}.}
  \label{fig:spec_xmm_ib_za07}
\end{figure*}

\begin{table}
\noindent
\centering
\caption{The best-fit parameters of the sources A and B spectra obtained with \xmm\ \mos\ and \intl\ in the 0.8--10 and 17--120~keV energy bands, respectively. Model: $TBabs*ZA07$. $C$ is the cross-calibration constant between the \xmm\ and \ib\ data points, while $C_{\rm XMM}$ is the cross-normalization constant between the source A and source B data.}
\label{tab:xmm_ibis_za07}
\vspace{1mm}
\begin{tabular}{c|c|c|c}
\hline
Parameter                   & Units                                        & Equation~(\ref{eq:za1}) & Equation~(\ref{eq:za2})      \\
\hline \hline
$N_{\rm H}$                 & $10^{22}$~cm$^2$                             & $1.022\pm0.008$         & $1.031\pm0.008$ \\
$\epsilon_0$                & keV                                          & $1.13\pm0.06$           & $1.73\pm0.11$   \\
$N_{\rm @keV}$              & $10^{-3}~ph~s^{-1}~cm^{-2}$                  & $21.6\pm0.2$            & $21.9\pm0.2$    \\
$C_{\rm XMM}$               &                                              & $0.858$ (fixed)         & $0.858$ (fixed) \\
$C$                         &                                              & $1.26\pm0.16$           & $1.05\pm0.13$   \\
$\chi^2_{\rm r}$/d.o.f.     &                                              & $1.19/1145$             & $1.17/1145$     \\
\hline
\end{tabular}\\
\vspace{3mm}
\end{table}

The above results qualitatively agree with \cite{Tsuji_2019}, who for the {\it cutoffpl} model obtained the cutoff energy at $18.8^{+4.2}_{-3.0}~^{+ 2.6}_{-2.1}$~keV (the first and second errors correspond to the statistic and systematic errors), which is close to our adopted value of 17~keV. In addition, \cite{Tsuji_2019} found that the \nus\ spectrum of \rx\ in the 3--20~keV band, is well described by the power-law model with the photon index of $\Gamma = 2.55\pm0.04~\pm0.02$, which agrees with our conclusion on the steepening of the power law slope in the hard X-ray band. 

\begin{figure*}
\includegraphics[width=0.49\textwidth]{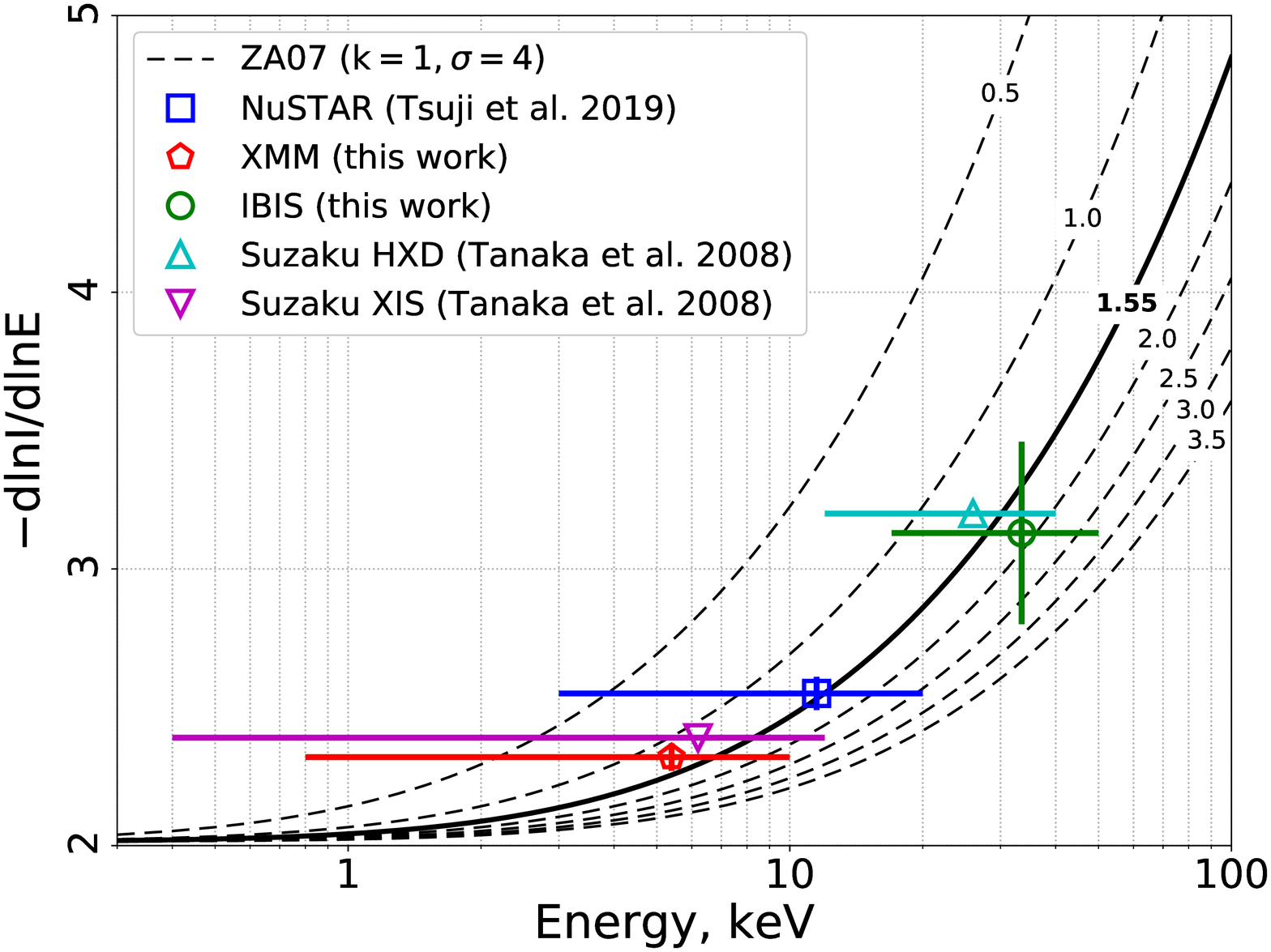}
\includegraphics[width=0.49\textwidth]{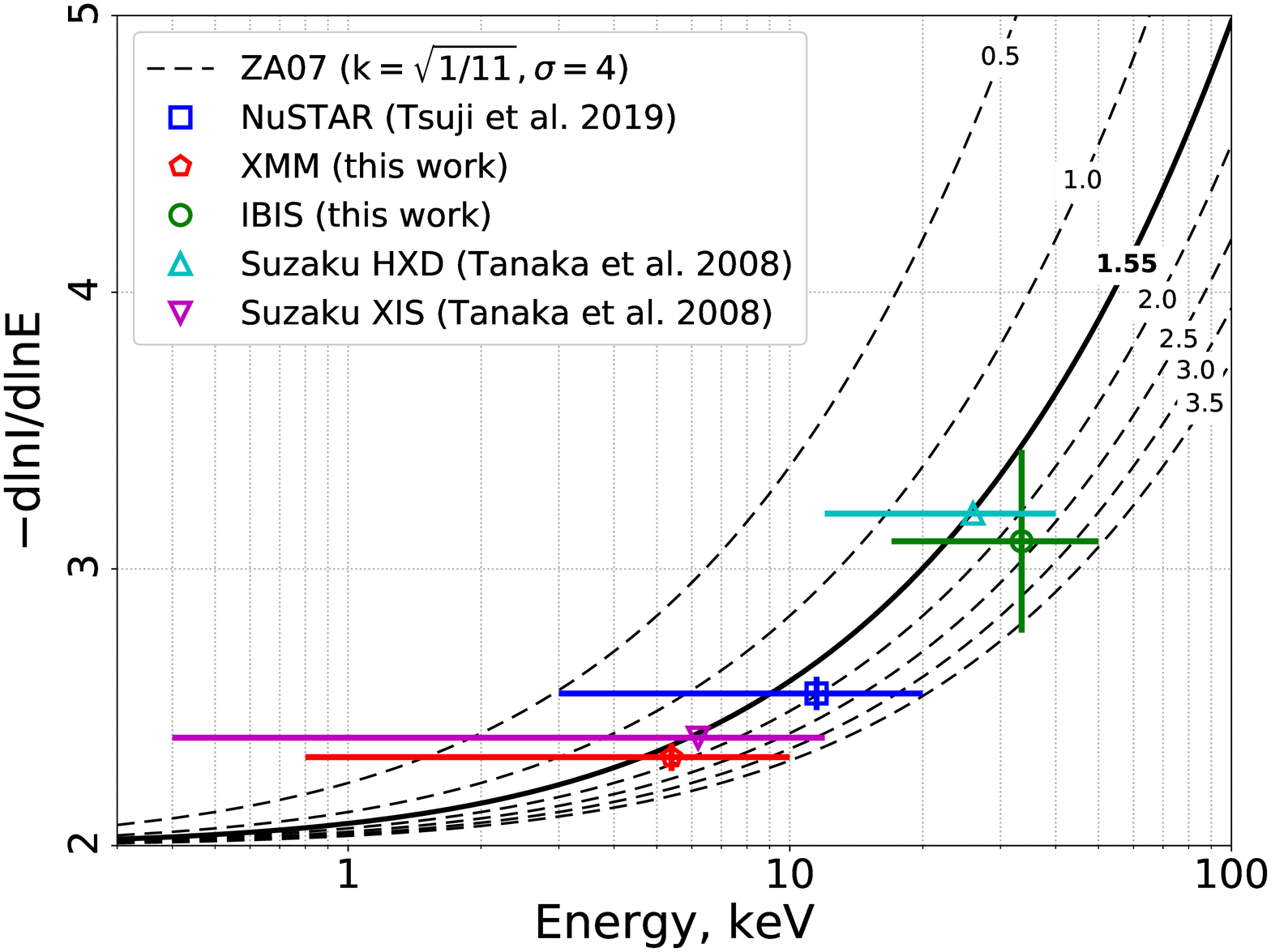}
  \caption{Power-law photon index values of \rx\ measured by different instruments. Dashed lines show a logarithmic derivative of ZA07 model for the $k=1$ (left) and $k=\sqrt{1/11}$ (right) as a function of energy $\epsilon$. Solid lines correspond to ZA07 model with $\epsilon_0=1.55\,{\rm keV}$ estimated for the Bohm limit regime with the shock velocity of $u = 3900~{\rm km\,s^{-1}}$ \citep{Tsuji_2019}. Different curves show the cases with different $\epsilon_0$ parameter in keV units.}
  \label{fig:phot_ind}
\end{figure*}

As regards physically motivated spectral models, we use the model of \cite{2007A&A...465..695Z} (hereafter ZA07), who gave an analytic approximation of the synchrotron emission of electrons accelerated by a non-relativistic shock-wave in a shell-type SNR, assuming that synchrotron losses dominate. Two flavors of this model are proposed by ZA07:
\begin{equation}
\frac{dN}{d\epsilon}\sim \epsilon^{-2} \left[1+0.46\left(\frac{\epsilon}{\epsilon_0}\right)^{0.6}\right]^{11/4.8}e^{-\sqrt{\epsilon/\epsilon_0}},
\label{eq:za1}
\end{equation}
and
\begin{equation}
\frac{dN}{d\epsilon}\sim \epsilon^{-2} \left[1+0.38\left(\frac{\epsilon}{\epsilon_0}\right)^{0.5}\right]^{11/4}e^{-\sqrt{\epsilon/\epsilon_0}},
\label{eq:za2}
\end{equation}
where $\frac{dN}{d\epsilon}$ is the observed spectrum in units of ${\rm phot\,s^{-1}\,cm^{-2}\,keV^{-1}}$ and $\epsilon$ is the photon energy. 
The former expression corresponds to the case when the ratio $k$ of the magnetic field strength upstream to downstream  $k=B_1/B_2=1$ (magnetic field is perpendicular to the shock front), while the latter is for $k=\sqrt{1/11}$ (isotropic magnetic field). In both cases, the compression factor $\sigma=4$ is assumed \citep[see][for details]{2007A&A...465..695Z}. The parameter $\epsilon_0$ is set by a competition of the synchrotron losses and the particle energy gain due to acceleration at the shock front, and depends, in particular, on the shock velocity and on the departures of the particle diffusion coefficient from the Bohm value. It does not correspond to an apparent cutoff or a break energy in the observed spectrum, because the power-law term with a positive index effectively compensates the slow exponential term at $\epsilon\sim \epsilon_0$ energies. Accordingly, the break/cutoff shows up at  higher energies than $\epsilon_0$. 

We fitted the broad-band spectrum of \rx\ with ZA07 model adding it as {\sc XSPEC} table model (see Fig.~\ref{fig:spec_xmm_ib_za07}). The model is in a good agreement with the \xmm\ \mos\ and \intl\ data for both cases of $k$ (see Table~\ref{tab:xmm_ibis_za07}). The estimations of $\epsilon_0$ at $1.13\pm0.06$~keV and $1.73\pm0.11$~keV for Eqs.~(\ref{eq:za1}) and {(\ref{eq:za2}), respectively, are consistent with \cite{Tsuji_2019}.

To avoid the normalization uncertainties of the \rx\ spectra obtained with different telescopes, we plot individual slopes as a function of energy. To this end, we used the  photon indexes of \rx\ measured by \nus\ \citep{Tsuji_2019} and {\it Suzaku}/{\sc XIS}/{\sc HXD} \citep{2008ApJ...685..988T} along with the \xmm\ and {\it INTEGRAL} photon indexes measured in this work. In Fig.~\ref{fig:phot_ind}, we compare the running photon index for the two ZA07 models given by Eqs.~(\ref{eq:za1}) and (\ref{eq:za2}) with the available observational data. The running photon index is evaluated as $\displaystyle \Gamma(\epsilon)=-\frac{d\ln N}{d\ln\epsilon}$ and shown in  Fig.~\ref{fig:phot_ind} with dashed lines, corresponding to different values of $\epsilon_0$. It is clear that qualitatively the progressive steepening of the spectrum is consistent with the ZA07 models, although uncertainties in observational data are substantial. In terms of the parameter $\epsilon_0$, both models suggest comparable values $\epsilon_0\sim1.0-2.0\,{\rm keV}$ and $\epsilon_0\sim1.5-2.5\,{\rm keV}$ for Eqs.~(\ref{eq:za1}) and (\ref{eq:za2}), respectively, which are consistent with values obtained from fitting procedure (see Table~\ref{tab:xmm_ibis_za07}). As expected, the model corresponding to $k=\sqrt{1/11}$ (see Eq.~(\ref{eq:za2})), prefers slightly higher values of $\epsilon_0$ than the model with $k=1$. For the estimated shock velocity in \rx\ $u=3900~{\rm km\,s^{-1}}$ and the diffusion coefficient close to the Bohm limit, the value of $\epsilon_0$ predicted by ZA07 model is $1.55\,{\rm keV}$ \citep{Tsuji_2019}. Therefore, the \ib\ data up to 50~keV in combination with the data of other observatories are consistent with the assumption that the shock in \rx\ operates in a regime close to the Bohm limit.

\section{Summary}
\label{sect:summary}

In this paper, we presented the first detailed study of \rx\ with {\it INTEGRAL/IBIS/ISGRI} in the hard X-ray energy band. The images of \rx\ obtained in the 17--27, 27--36, 36--50, and 17--60~keV bands are in a good agreement with the more detailed 1--10~keV \xmm\ map of \rx, which points to a single emission mechanism operating in the soft and hard X-ray bands. The hard X-ray {\it IBIS/ISGRI} image of \rx\ is dominated by two extended X-ray sources spatially coincident with the brightest parts of the SNR. Considering the shell structure of \rx, we find a good agreement between the position of the shocks in the hard and soft X-ray bands.

The spectral analysis of the \ib\ data shows that in the 17--120~keV energy band, the \rx\ spectrum is well described by a power-law model with $\Gamma=3.13^{+0.36}_{-0.33}$, which is significantly steeper than $\Gamma=2.32\pm0.05$ determined from \xmm/\mos\ data in the 1--10~keV band. The difference in indexes points toward a change in the slope of the power-law spectrum during the transition from the soft to the hard X-ray bands.

Simultaneous fitting of the \xmm\ \mos\ and \ib\ data reveals that \ib\ data are better described by a broken power-law model than by an exponential cutoff model, if the break and cutoff energies are fixed to 17~keV for both models. Obtaining better constraints is difficult, since we are dealing with the complex diffuse source and absolute normalizations of the \xmm\ and \ib\ spectra for extended regions are hard to get, especially for the coded-mask telescopes.

Nevertheless, the $d\ln I/d\ln\epsilon$ diagram, which is free from the normalization issues shows that the photon index estimate based on the \ib\ data is not dissimilar from the predictions of the \cite{2007A&A...465..695Z} analytical model of the synchrotron-photon spectrum of electrons accelerated by non-relativistic shock-wave in a young shell-type SNR. We conclude that this model well describes the observational data with $\epsilon_0$ energy $\sim1-2$~keV, in agreement with expectations for the acceleration regime close to the Bohm limit.

\section*{Acknowledgments}
This work is based on observations with INTEGRAL, an ESA project with instruments and the science data centre funded by ESA member states (especially the PI countries: Denmark, France, Germany, Italy, Switzerland, Spain), and Poland, and with the participation of Russia and the USA. This work is supported by the Russian Science Foundation (grant 19-12-00369).

\bibliographystyle{mnras}
\bibliography{references}

\bsp	% typesetting comment
\label{lastpage}
\end{document}